\documentstyle[aps,graphicx,prl,multicol]{revtex}
\input{epsf.tex}
\begin{document}
\draft
\title{The low temperature Fulde-Ferrell-Larkin-Ovchinnikov phases in 
3 dimensions}
\author{R. Combescot and C. Mora }
\address{Laboratoire de Physique Statistique,
 Ecole Normale Sup\'erieure*,
24 rue Lhomond, 75231 Paris Cedex 05, France}
\date{Received \today}
\maketitle

\begin{abstract}
We consider the nature of the Fulde-Ferrell-Larkin-Ovchinnikov 
(FFLO) phases in three dimensions at low temperature. We introduce a 
new method to handle the quasiclassical equations for superconductors 
with space dependent order parameter, which makes use of a Fourier 
expansion. This allows us to show that, at $T=0$, an order parameter 
given by the linear combination of three cosines oscillating in 
orthogonal directions is preferred over the standard single cosine 
solution. The transition from the normal state to this phase is first order, 
and quite generally the transition below the tricritical point to the FFLO 
phases is always first order.
\end{abstract}

\begin{multicols}{2}

Although they have been introduced quite a long time ago 
\cite{ff,larkov}, Fulde-Ferrell-Larkin-Ovchinnikov (FFLO) phases are 
still of very high interest. On the experimental side they should appear 
as the superconducting phases in extremely high field superconductors, 
which are obviously of very high practical interest. On the theoretical 
side, quite surprisingly, a precise understanding of the order paramater 
and nature of the transitions in these phases has not yet been reached, 
and this question is naturally of essential importance for the 
experimental identification of these phases, which have not been 
observed to date unambiguously. These FFLO phases correspond to a 
spontaneous symmetry breaking of the standard BCS superfluid phase 
in the presence of an effective field, which has the only effect of 
inducing a difference in chemical potential between the two populations 
involved in the formation of Cooper pairs. This leads to an 
inhomogeneous superfluid with a space dependent order parameter. The 
chemical potential difference is due for example to the coupling of the 
effective field with the spins of the particles. Naturally in standard 
superconductors a real magnetic field gives rise to a coupling to the 
electronic orbital degrees of freedom which is much stronger than the 
coupling to the spins and is ordinarily responsible for the critical field. 
Here we investigate the original problem considered by Larkin and 
Ovchinnikov where this orbital coupling is absent or essentially 
negligible. Accordingly we consider the simplest model with a spherical 
Fermi surface.

In their original work Larkin and Ovchinnikov \cite{larkov} (LO) 
explored at $T = 0$ the possibility of a second order phase transition 
and looked, in the three-dimensional case, for the order parameter 
corresponding to the lowest energy. They found that the stablest state 
has a simple one-dimensional space dependence $ \Delta ({\bf  r}) \sim 
\cos ({\bf  q}.{\bf  r}) $ for the order parameter. Nevertheless they left 
open the possibility that a first order transition could exist at higher 
fields. However looking for a first order transition is a much more 
complicated problem than exploring a second order one. Indeed LO 
could solve for the second order transition by performing an expansion 
of the free energy up to fourth order in powers of the order parameter, 
while no expansion is in principle allowed in the case of a first order 
transition. Similarly LO could restrict the space dependence to a sum of 
plane waves, while for first order this is no longer correct. In this paper 
we consider the original problem raised by LO and show, by solving 
the quasiclassical equations with the use of a Fourier expansion, that the 
transition is always first order. We find in particular that, at $T=0$, the 
preferred order parameter has a cubic symmetry and is surprisingly very 
near one of the order parameters investigated explicitely by LO.

The question of the order of the transition to the FFLO phases has given 
rise recently to a good deal of work. Burkhardt and Rainer \cite{br} 
used quasiclassical equations to show that, for the two-dimensional 
case, the transition to $ \Delta ({\bf  r}) \sim \cos ({\bf  q}.{\bf  r}) $ 
stays second order all the way up to the tricritical point, where the 
FFLO transition line meets the standard BCS line. Very recently we 
have investigated \cite{mc} the specific nature of the resulting order 
parameter in the low temperature regime. On the other hand in the three-
dimensional case, Matsuo \cite{matsuo} \emph{et al} showed, again by 
solving quasiclassical equations, that the transition to this same order 
parameter becomes first order at low temperature, above $ T/ T_{c0}= 
0.075$, where $T_{c0}$ is the standard critical temperature for equal 
spin chemical potentials. It stays then first order up to the tricritical 
point. A different line of attack has been to investigate this problem in 
the vicinity of this tricritical point (located at $T_{\rm{tcp}} / T_{c0} = 
0.561 $). Indeed because of the proximity of the second order transition 
to the standard BCS phase with uniform order parameter, an expansion 
up to sixth order \cite{buz2,buz1,cm} in the order parameter can 
always be used. This has been done by Houzet \emph{et al} 
\cite{buz1} who examined a number of possible order parameters and 
concluded that the best one was indeed the simple cosine $ \Delta ({\bf  
r}) \sim \cos ({\bf  q}.{\bf  r}) $. Quite recently we have investigated 
this same problem analytically \cite{cm} and came to the same 
conclusion by showing that the states with the smallest number of plane 
waves are favored.

These explorations brought a somewhat surprising information. In the 
vicinity of the tricritical point \cite{buz1,cm} the location of the first 
order transition is extremely close to the standard FFLO second order 
transition. This remains true \cite{matsuo}, for the transition to the 
simple cosine, down to the lowest temperatures. This feature suggests 
that the first order transition is not very far from being second order. If 
this were the case, this would allow an expansion in powers of the 
order parameter (for example up to sixth order) to remain valid at any 
temperature, and not only near the tricritical point. This has been 
explored by Houzet \emph{et al} \cite{buz1}, but this approach does 
not work because somewhat below the tricritical point the coefficient of 
the sixth order term changes sign. Actually one can show that, whatever 
the order of the expansion, the coefficients are changing sign when the 
temperature is lowered. So an expansion in powers of the order 
parameter can not succeed. This leads to the conclusion that the 
quasiclassical approach is necessary to deal with this problem. On the 
other hand another feature valid near the tricritical point can be used. 
Indeed our exploration \cite{cm} showed that the order parameter at the 
first order transition is very near the simple plane waves superposition 
coming in the LO analysis of the second order phase transition. Hence 
we may expect that a Fourier expansion will be a good approach at any 
temperature. In this paper we show indeed that introducing a Fourier 
expansion in the quasiclassical equations is a very efficient scheme to 
deal with this non-linear problem. We will see that such an expansion 
converges very rapidly toward the exact result and that, as a 
consequence, the first few terms give an excellent approximation. 

Here we will use Eilenberger's original equations which are enough to 
handle our problem. However an essential strength of the quasiclassical 
method is its ability to be generalized for full inclusion of many body 
effects \cite{sererain}. The quasiclassical Green's functions $ g ( 
\omega , \hat{ {\bf  k} }, {\bf  r}) =  \frac{i}{ \pi } \int d \xi _{k} G( 
\omega , {\bf  k}, {\bf  r}) $ are obtained by " $ \xi $-integrating " the 
usual temperature Green's functions $ G(\omega _{n} , {\bf  k}, {\bf  
r}) $. Here $\xi _{ {\bf  k}}$ is the kinetic energy measured from the 
average Fermi level $ (1/2) ( \mu _{\uparrow}+ \mu_{\downarrow})$. 
The frequency $ \omega $ will turn at some stage to be the Matsubara 
frequency $ \omega _{n} = \pi T (2n+1) $. The existence of the FFLO 
phases is due to the effective field $ \bar{\mu } =  (\mu _{\uparrow} - 
\mu_{\downarrow})/ 2 $ produced by the difference between spin up 
and spin down chemical potentials. It will merely appear \cite{larkov} 
by the replacement of $ \omega _{n}$ by $ \bar{\omega  }_{n} = 
\omega _{n} - i \bar{\mu } $ in the calculations. We simplify the 
writing by taking $ \hbar = 1$ and $ m = 1/2 $. Eilenberger's equations 
for the off-diagonal $ f ( \omega , \hat{ {\bf  k} }, {\bf  r})$ 
quasiclassical propagators then read \cite{eil1} :
\begin{eqnarray}
( \omega + {\bf  k}.{\bf \nabla}) f ( \omega , \hat{ {\bf  k} }, {\bf  r}) 
= \Delta ( {\bf  r}) 
g ( \omega , \hat{ {\bf  k} }, {\bf  r}) \nonumber  \\
( \omega - {\bf  k}.{\bf \nabla}) f ^{+} ( \omega , \hat{ {\bf  k} }, {\bf  
r}) = \Delta ^{*}( {\bf  r}) 
g ( \omega , \hat{ {\bf  k} }, {\bf  r})
\label{eq1}
\end{eqnarray}
The diagonal propagator $g$ is given in terms of $f$ and $f ^{+}$ by 
the normalization condition :
\begin{eqnarray}
g ( \omega , \hat{ {\bf  k} }, {\bf  r}) = (1 - f ( \omega , \hat{ {\bf  k} 
}, {\bf  r}) 
f ^{+} ( \omega , \hat{ {\bf  k} }, {\bf  r}))^{1/2}
\label{eq2}
\end{eqnarray}
from which one can \cite{eil1} deduce easily an equation for $g$ 
similar to Eq.(1). Actually the equations Eq.(1) for $f$ and $f ^{+}$ 
are related since \cite{eil1} $f  ^{*} (- \omega , \hat{ {\bf  k} }, {\bf  
r}) = f  ^{*} (\omega , - \hat{ {\bf  k} }, {\bf  r}) = f ^{+} ( \omega , 
\hat{ {\bf  k} }, {\bf  r}) $ with analogous relations holding for $ g ( 
\omega , \hat{ {\bf  k} }, {\bf  r})$.

We will consider here the very wide class of order parameters which 
admit a three-dimensional Fourier expansion. The order parameters 
considered by LO fall in particular in this class. We will also assume 
that the order parameter is real. It has indeed been shown \cite{cm} that 
this class is favored near the tricritical point, and the order parameters 
considered specifically by LO are also real. For clarity we will only 
consider explicitely \cite{longpaper} the case of a simple cosine order 
parameter $\Delta (x) = 2 \Delta \cos (qx)$ (the value of $q$ and $ 
\Delta $ are ultimately obtained by minimization, see below) and then 
explain the modifications occuring when one goes to the general case. 
For fixed $ {\bf  k}$ Eilenberger's equations are a set of first order 
equations for the variation of the Green's functions along $ {\bf  k}$. 
So we take a reduced variable along this direction by setting $ {\bf  r} = 
{\bf  k} X $ which gives $\Delta (x) =  2 \Delta \cos (QX)$ where we 
have introduced $ Q = k _{F} q \cos \theta $ with $ \theta $ the angle 
between $ {\bf  k}$ and the $x$ axis. Then we make a Fourier 
expansion of the Green's functions :
\begin{eqnarray}
f(X) = \sum_{n} f _{n} \: e ^{ inQX}
\label{eq3}
\end{eqnarray}
and similarly for $ f ^{+}(X)$ and $g(X)$. Parity implies $ f(-X) = f 
^{+}(X)$ and $ g(-X) = g(X)$, which gives $ f ^{+}_{-n} = f _{n}$ 
and $ g _{-n} = g_{n}$. Inserting these expansions in Eq.(1) we find 
the recursion relations:
\begin{eqnarray}
d _{n} = - \frac{nQ \Delta }{ \omega ^{2}+ (nQ) ^{2}} (g _{n-1}+ g 
_{n+1})  \nonumber  \\
g _{n} = \frac{ \Delta }{nQ} (d _{n-1}+ d _{n+1})
\label{eq4}
\end{eqnarray}
where we have set $ d _{n} = (f _{n} - f ^{+}_{n})/2i  $, from which 
$ f _{n} = (i - \omega /nQ ) d _{n}$ is obtained. These equations show 
that $ g _{2p+1}= 0$ and $ d _{2p}= 0$. Eq.(4) are linear and must be 
supplemented by the normalization condition Eq.(2). The $n=0$ 
component is enough and it provides us precisely with the spatial 
integral $ g_{0} =  \int \, d{\bf  r} \: g  ( \omega , \hat{ {\bf  k} }, {\bf  
r})$ needed to calculate the free energy :
\begin{eqnarray}
g ^{2}_{0} = 1 - \sum_{n \neq 0} (g_{n} g_{-n} + f _{n} f  ^{+}_{-
n})
\label{eq5}
\end{eqnarray}
The critical temperature $ T $ is then obtained by writing the equality  
$\Omega _{s} = \Omega _{n}$ between normal and superfluid state 
free energies. When this is done from Eilenberger's expression 
\cite{eil} for $\Omega _{s}$, we find for our simple cosine, after some 
transformations \cite{longpaper}, that  $ \ln [T / T _{sp}(\bar{\mu 
}/T)]$ is the minimum of:
\begin{eqnarray}
- \, \frac{2 \pi  T }{\Delta ^{2}} 
\sum_{n=0}^{ \infty} \int_{\omega _{n}}^{ \infty}
d \omega  {\mathrm{Re}} [ \int_{0}^{1} \, du g _{0} (\bar{\omega } , 
\bar{q}u )
- 1 + \frac{ \Delta ^{2}}{ \bar{\omega } ^{2}}] 
\label{eq6}
\end{eqnarray}
with $ \bar{\omega } = \omega - i \bar{\mu } $ and where $ T 
_{sp}(\bar{\mu }/T)$ is the location of the spinodal transition, which is 
the simple continuation of the standard BCS transition. Since for 
homogeneity the above quantity is only a function of $ T / \bar{\mu }$, 
$ \Delta / \bar{\mu }$ and $ \bar{q} / \bar{\mu } $ one has to minimize 
with respect to $ \Delta / \bar{\mu }$ and $ \bar{q} / \bar{\mu } $, at 
fixed $ T / \bar{\mu }$. In this way one finds the critical temperature 
$T$ as a function of $ T / \bar{\mu }$, hence $T$ as a function of 
$\bar{\mu }$.

Now the interesting point is the large $n$ behaviour of $ g _{n}$ and $ 
d _{n}$. The general solution of the recursion relations Eq.(4) is a 
linear combination of two independent solutions. For large $n$ the 
recursion relation for $g _{n}$ simplifies into $ \Delta  ^{2}( g _{n+2} 
+ g _{n-2}) +  (nQ) ^{2} g _{n} = 0 $. This equation has very rapidly 
growing solutions behaving as $ g _{2p+2} \sim (-1) ^{p} (2Q/ \Delta 
)^{2p} (p!) ^{2}$. Naturally these solutions are not physically 
acceptable. On the other hand the recursion relation has also a solution 
satisfying $ g _{n+2} \ll  g _{n} \ll  g _{n-2} $ and behaving as $ g 
_{2p} \sim (-1) ^{p} (\Delta / 2Q)^{2p} (1/p!) ^{2}$, which is the 
physical solution we are looking for. This solution is found only if $ g 
_{0}$ and $ g _{2}$ are related by a specific boundary condition.

Up to now no approximation has been made. Now the very fast 
decrease of $ g _{n}$ and $ d _{n}$ provides an easy way to obtain a 
set of approximate solutions, which moreover converges rapidly to the 
exact one, all the more since these are $ g _{n} ^{2}$ and $ d _{n} 
^{2}$ which come in Eq.(5) for the calculation of $ g _{0}$. Since  $ 
g _{n}$ and $ d _{n}$ are very small for large $n$ we just take $ g 
_{n}$ and $ d _{n+1}$to be zero for some fixed value $n=N+1$ and 
beyond. This serves as boundary condition. Then we work backward 
to obtain the whole set of Fourier components and normalize them 
properly through the normalization condition Eq.(5). When we let $ N 
\rightarrow \infty $ we find the exact result for $g_{0}$. The recursion 
relations Eq.(4) are very convenient and very fast for a numerical 
implementation and in practice the situation is not very different from 
having an analytical expression for $ g_{0}$. Basically the solution of 
Eq.(1) is reduced to some simple algebra. The simplest of the set of 
converging approximations corresponds to take $N = 1$ and it is given 
explicitely by :
\begin{eqnarray}
g_{0} = [ 1 + 2 \Delta ^{2} \frac{ \omega ^{2} - Q ^{2}}{( \omega 
^{2} + Q ^{2})^{2}} ] ^{- \frac{1}{2} }
\label{eq7}
\end{eqnarray}
This is already a quite non trivial approximation. Since it is correct up to 
order $  \Delta ^{2}$ it gives the proper location for the standard FFLO 
second order transition line. Moreover it gives qualitatively and 
semiquantitatively the correct results, with a first order transition down 
from the TCP which becomes a second order transition at low 
temperature in agreement with Matsuo \emph{et al} \cite{matsuo}. The 
switch from first to second order occurs at $T/ T_{c0}=0.195$ to be 
compared with the exact result $ 0.076$. When we go to $N=3$ we 
find $0.063$ and for $N=5$ our results coincide with those of Matsuo 
\emph{et al} at $0.076$.

Let us now consider the various steps in extending this method to full 
generality \cite{longpaper}. First we want to include additional Fourier 
components of the order parameter. This is done for example by 
solving first the problem for the lowest Fourier component, then 
inserting the result in the self-consistency equation to obtain a new 
order parameter and iterating this process until it is converged. This 
process can be carried out for a given number of Fourier components, 
by projecting out the unwanted ones. Naturally for a fixed number of 
Fourier components in the order parameter one has to generalize the 
recursion relations Eq.(4), but they still have the property that, for the 
physical solution, the Fourier components have a factorial decrease for 
large order. This makes again possible to set them to zero beyond some 
order, and to solve in this way for the other components. The system of 
equations one has to deal with is slightly less convenient than Eq.(4), 
but this is a linear system which is easily dealt with by standard 
numerical methods. In practice, for our specific problem, we have 
found that, at the transition from normal state to FFLO phases, the 
coefficient $ \Delta _{1}$ of the next harmonic $ \cos(3qx)$ is of order 
$10 ^{-2}$ compared to the first one and produces a very small 
correction. This is in agreement with our findings \cite{cm} near the 
tricritical point. We have found this result for order parameters having 
either a single cosine or two cosines. We assumed it for order 
parameters with three and four cosines that we have considered.

Next we can take the order parameter to be any linear combination of 
cosines, with wavevectors in different directions. More generally we 
could take a three-dimensional Fourier expansion for the order 
parameter. Naturally in practice this becomes too heavy for a large 
number of components. This generalization amounts to replace our 
above index $n$ in the Fourier expansion Eq.(3) by a set of integers, 
for example $(n_1,n_2,n_3)$ for the order parameter $\Delta ( {\bf  r}) 
= 2 \Delta_1 \cos ({\bf  q} _{1}.{\bf  r}) + 2 \Delta_2 \cos ({\bf  q} 
_{2}.{\bf  r}) +2 \Delta_3 \cos ({\bf  q} _{3}.{\bf  r})$. This means 
that our Fourier components are now indexed by a set of points in a 
three-dimensional space. We make now our cut-off for $ | n_1| +  | n_2| 
+ | n_3| \ge N $. Naturally we have much more components but our 
linear algebraic treatment goes essentially in the same way and it stays 
still quite manageable. We note in particular that the values of the ${\bf  
q} _{i}$'s do not need to have any simple relation (in particular they 
are not necessarily orthogonal).

An important point, which helps indeed very much the numerical 
calculation, is that the Fourier expansion converges quite rapidly as we 
already mentionned for the simple cosine. We have found that, for the 
case of two cosines, $N=3$ was already quite enough and that $N=5$ 
did not give any sizeable change. Actually, except for a single cosine, 
the results for $N=1$ are already quite reasonable quantitatively.

In this way we have been able to make some exploration of the 
evolution of the critical temperature when the relative weights $ \Delta 
_{i}$ of the cosines and the corresponding wavevectors ${\bf  q} 
_{i}$ vary. For example, in the case of two cosines, we have kept the 
wavevectors orthogonal or with an angle $ \pi /3 $ but minimized 
independently the two weights. We have found that, in either case, the 
optimum is for equal weights. Conversely we have varied continuously 
the angle between the two directions, keeping the weights equal. 
Similarly for three cosines, taking equal weights, we have kept the 
directions in rhomboedral symmetry, but allowed the rhomboedral 
angle to vary. These investigations make quite likely that, both for the 
cases of two and of three cosines, the stable phase is obtained for equal 
weights $ \Delta _{i}$ and for orthogonal wavevectors ${\bf  q} _{i}$ 
with same lengths. In the following we present results only for these 
simple cases.
\begin{figure}[h]
\begin{center}
\includegraphics[scale=0.65]{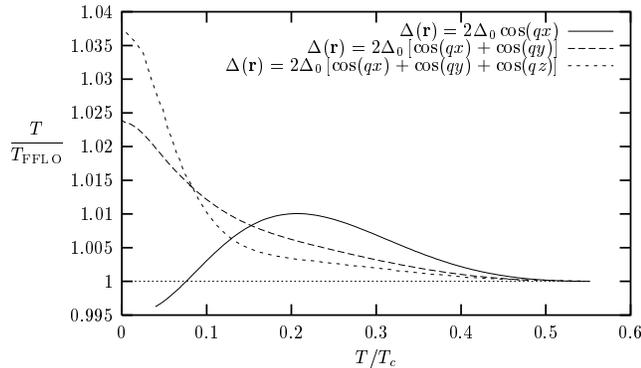}
\caption{Temperature $T$ of the first order transition compared to the 
critical temperature $T _{FFLO}$ for the standard FFLO second order 
transition, as a function of the ratio $T / \bar{ \mu } $. This is shown 
for various order parameters with one, two or three cosines oscillating 
in orthogonal directions. Actually we give on the horizontal axis the 
ratio $ T / T_c$, which is an increasing function of $T / \bar{ \mu }$ 
and is more physical. Note that, for low temperature, it is essentially 
equivalent to consider that this is $ \bar{ \mu }/\bar{\mu} _{FFLO}$ 
which is given on the vertical axis.}
\label{fonctT}
\end{center}
\end{figure}
The comparison between our results for order parameters with one, two 
and three cosines can be seen in Fig.1. Naturally the phase which is 
physically present is the one which has the highest critical temperature $ 
T / T _{FFLO}$, where $T _{FFLO}$ is the critical temperature for 
the standard FFLO second order transition. For $ T / T _{c} > 0.154 $ 
the phase with a single cosine is the stable one, in agreement with 
previous results near the tricritical point \cite{buz1,cm}. On the other 
hand in the range $ 0.080 < T / T _{c} < 0.154 $, the stable order 
parameter has two cosines. Finally the phase with the three cosines is 
the stable one below $ 0.080 T _{c}$. As a result the transition from 
the normal state to the stable FFLO phase is always a first order 
transition in three dimensions, in contrast with the original proposal 
\cite{larkov} of LO. We note that the succession of transitions we have 
found is in agreement with the scenario proposed by Houzet \emph{et 
al} \cite{buz1}. It is by no means obvious that increasing the number of 
cosines does not lead to even further increase in stability. We have 
looked in this direction by considering an order parameter with four 
cosines, with symmetrical directions pointing toward the corners of a 
cube (the angle between any two directions is $70.5^{\circ}$; note that 
this is an aperiodic order parameter). The result for $ T / T _{FFLO}$  
($1.016$ at $T=0$) is worst than for three cosines or even for two 
cosines. Qualitatively this goes in the direction proposed recently 
\cite{bowers} of an effective repulsion between the directions of the 
various wavevectors involved in the order parameter, with an angle of 
repulsion of order $ 2 \arccos(1/ \bar{q}) = 67^{\circ}$ for $ \bar{q} = 
1.2$ . Taking into account that $ \Delta $ is not zero pushes this angle 
somewhat above $67^{\circ}$ and this agrees qualitatively with the fact 
that we can fit three, but not four directions on the unit sphere.

It is remarkable and unexplained that $T/T_{FFLO}$ stays so near 1 
for all our results, whatever the choice of the order parameter. However 
since in all cases we find that at the transition $ \Delta_0 / \bar{ \mu }$ 
is of order $0.2$, the order parameter stays rather small. This seems to 
imply that the transition is not far from being second order, which is 
consistent with the fact that $T/T_{FFLO}$ stays near $1$. It would be 
interesting to explore if this feature is a mere coincidence by going to 
more complicated situations, for example by taking into account Fermi 
liquid effects \cite{br} within the framework of Landau's theory.

\noindent
* Laboratoire associ\'e au Centre National
de la Recherche Scientifique et aux Universit\'es Paris 6 et Paris 7.

\end{multicols}

\begin{references}
\bibitem{ff}P. Fulde and R. A. Ferrell, Phys.Rev. {\bf 135}, A550 
(1964).
\bibitem{larkov}A. I. Larkin and Y. N. Ovchinnikov, ZhETF {\bf 47}, 
1136 (1964) [Sov. Phys. JETP {\bf 20}, 762 (1965)].
\bibitem{br}H. Burkhardt and D. Rainer, Ann.Physik {\bf  3}, 181 
(1994).
\bibitem{mc} C. Mora and R. Combescot, cond-mat/0306575
\bibitem{matsuo}S. Matsuo, S. Higashitani, Y. Nagato and K. Nagai, 
J. Phys. Soc. Japan, {\bf  67} 280 (1998).
\bibitem{buz2}A. I. Buzdin and H. Kachkachi, Phys. Lett. A {\bf  
225} 341 (1997).
\bibitem{buz1}M. Houzet, Y. Meurdesoif, O. Coste and A. I. Buzdin, 
Physica C {\bf  316}, 89 (1999).
\bibitem{cm}R. Combescot and C. Mora, Eur.P.J. B {\bf 28}, 397 
(2002).
\bibitem{sererain}J. W. Serene and D. Rainer, Physics Reports {\bf  
101} 221 (1983).
\bibitem{eil1}G. Eilenberger, Z. Phys. {\bf  214} 195 (1968).
\bibitem{longpaper}Details will be published elsewhere.
\bibitem{eil}G. Eilenberger, Z. Phys. {\bf  182} 427 (1965).
\bibitem{bowers}J. A. Bowers and K. Rajagopal, Phys. Rev. D {\bf 
66} 065002 (2002).

\end{references}
\end{document}